\documentclass[aps, reprint, 10pt, superscriptaddress, twocolumn, prb]{revtex4-2}

\usepackage[dvipsnames]{xcolor}
\usepackage{csquotes}
\usepackage{graphicx}
\usepackage{dcolumn}
\usepackage{bm}
\setcitestyle{super}

\begin{document}

\preprint{APS/123-QED}

\title{Dynamical Disorder in the Mesophase Ferroelectric $\mathrm{HdabcoClO_4}$: \\ A Machine-Learned Force Field Study}

\author{Elin Dypvik Sødahl}
\affiliation{Department of Mechanical Engineering and Technology Management, Norwegian University of
Life Sciences, N-1433 Norway.}

\author{Jesús Carrete}
\affiliation{Instituto de Nanociencia y Materiales de Aragón (INMA), CSIC-Universidad de Zaragoza, ES-50009 Spain}

\author{Georg K. H. Madsen}
\affiliation{Institute of Materials Chemistry, TU Wien, A-1060 Austria}

\author{Kristian Berland}
\email[E-mail: ]{kristian.berland@nmbu.no}
\affiliation{Department of Mechanical Engineering and Technology Management, Norwegian University of
Life Sciences, N-1433 Norway.}

\date{\today}
\begin{abstract}
Hybrid molecular ferroelectrics with orientationally disordered mesophases offer significant promise as lead-free alternatives to traditional inorganic ferroelectrics owing to properties such as room temperature ferroelectricity, low-energy synthesis, malleability, and potential for multiaxial polarization.
The ferroelectric molecular salt $\mathrm{HdabcoClO_4}$
is of particular interest due to its ultrafast ferroelectric room-temperature switching. 
However, so far, there is limited understanding of the nature of dynamical disorder
arising in these compounds. 
Here, we employ the neural network \textsc{NeuralIL} to train a machine-learned force field (MLFF) with 
training data generated using density functional theory. 
The resulting MLFF-MD simulations exhibit phase transitions and thermal expansion in line with earlier reported experimental results, 
for both a low-temperature phase transition coinciding with the orientational disorder of $\mathrm{ClO_{4}^{-}}$ molecules and the onset of rotation of $\mathrm{Hdabco^{+}}$ and $\mathrm{ClO_{4}^{-}}$-molecules in a high-temperature phase transition. 
We also find proton transfer even in the low-temperature phase, which increases with temperature
and leads to associated proton disorder as well as the onset of disorder in the direction of the hydrogen-bonded chains. 
\end{abstract}

\maketitle

\section{\label{sec:intro} Introduction}

Hybrid molecular crystals and salts have recently attracted much interest 
due to their vast potential application range, including as electrolytes,\cite{pringle_recent_2013,armel_organic_2011,zhu_organic_2019} 
barocaloroics,\cite{salgado-beceiro_hybrid_2021} piezo-, and ferroelectrics.\cite{harada_plasticferroelectric_2021,das_harnessing_2020,shi_symmetry_2016,pan_recent_2021, walker_electromechanical_2024, walker_electric_2020} 
Moreover, the possibility of using room-temperature synthesis with low-energy methods, such as 3D-printing,\cite{jin_organic_2014} slow evaporation,\cite{owczarek_flexible_2016,deng_novel_2020,lan_cation_2021} and spin coating, \cite{harada_directionally_2016} allow for environmentally friendly production and flexible device integration. 
As different molecular species can be combined in many ways,
they offer immense design flexibility, which can circumvent the need for toxic molecules and/or scarce elements.
Some of these molecular crystals and salts, especially those consisting of globular (i.e., cage-like, disk-like, or cylindrical) \cite{das_harnessing_2020, zhu_organic_2019} molecules
can host plastic mesophases where the molecular species become orientationally disordered while retaining crystalline order.\cite{timmermans_plastic_1961} 
The onset of the orientational disorder, 
can also result in a marked increase in the number of
facile slip planes, contributing to the possibility of fusing or molding the molecular crystals into desired shapes and this class of materials are therefore often referred to as plastic crystals.\cite{das_harnessing_2020, mondal_metal-like_2020}

The degree of disorder can vary between plastic crystals, 
and some also display transitions between 
partly and fully orientationally disordered mesophases.\cite{harada_directionally_2016, olejniczak_pressuretemperature_2018, yoneya_molecular_2020}
The large entropy change in the transition from an ordered low-temperature phase to a disordered plastic phase\cite{li_colossal_2019, liu_giant_2022,aznar_reversible_2020,li_understanding_2020} 
can also be used for thermal storage and barocaloric cooling applications.\cite{lloveras_colossal_2019, salgado-beceiro_hybrid_2021}
These materials also have potential as effective ionic conductors. \cite{pringle_recent_2013,armel_organic_2011, zhu_organic_2019} For ferroelectric plastic crystals, 
the transition to the plastic mesophase often coincides with a transition to a paraelectric phase.\cite{harada_plasticferroelectric_2021,richard_jd_tilley_insulating_2013,harada_ferroelectricity_2018,harada_plasticferroelectric_2019, gonzalez-izquierdo_r--3-hydroxyquinuclidiumfecl4_2021} 

Molecular components of plastic crystals include neutral species such as paraffins and cycloalkanes, \cite{das_harnessing_2020} cationic species such as derivates of quinuclidium, dabco (1,4-diazabicyclo[2.2.2]octane), and tetramethylamine,\cite{shi_symmetry_2016, harada_plasticferroelectric_2021} and anionic species such as $\mathrm{ClO_{4}^{-}}$, $\mathrm{FeCl_{4}^{-}}$, $\mathrm{OCN^{-}}$, and $\mathrm{H_{2}PO_{4}^{-}}$.\cite{pringle_recent_2013, harada_plasticferroelectric_2021} 
Recently, we attempted to uncover 
novel ferroelectric molecular crystals from the 
Cambridge Structural Database \cite{seyedraofi_PT_CSD,dypvik_sodahl_ferroelectric_2023} 
finding 20 new systems that are likely to be both ferroelectric and plastic crystals.\cite{dypvik_sodahl_ferroelectric_2023}

\begin{figure}
    \centering
    \includegraphics[scale=0.4]{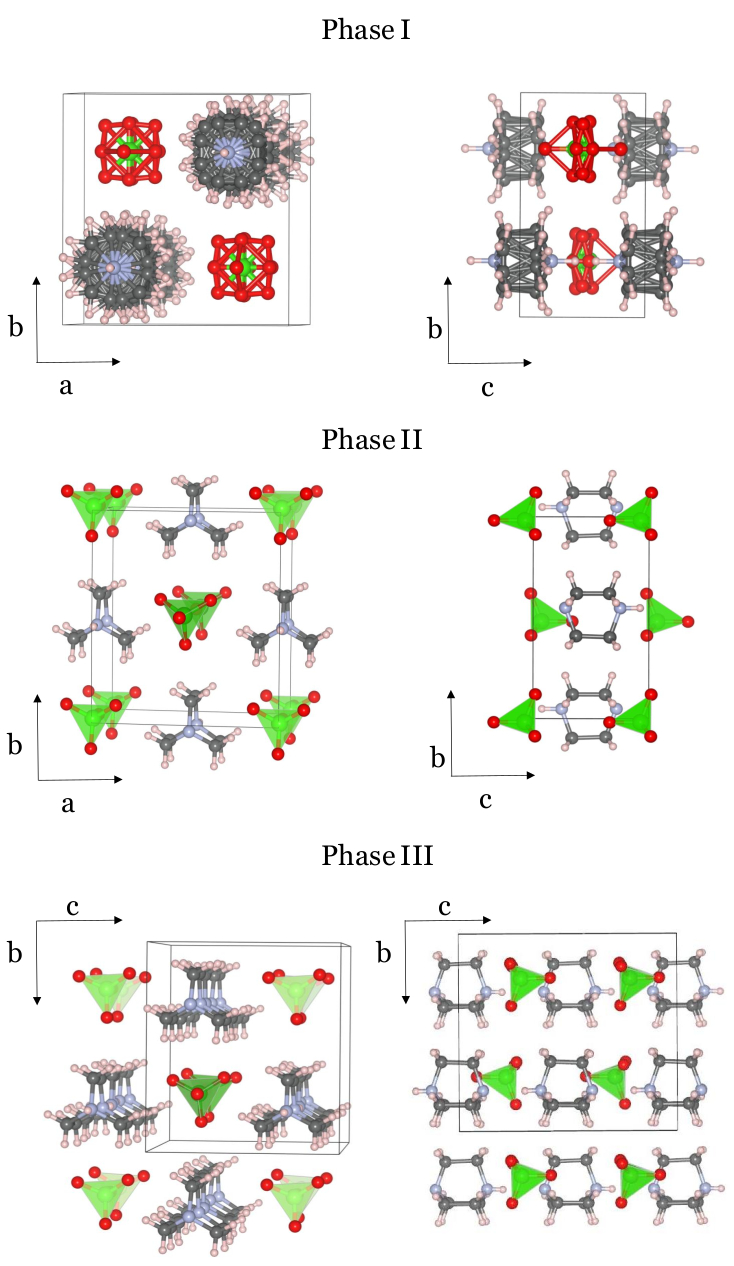}
    \caption{Illustration of the reported experimental crystal structures of phases I, II, and III of $\mathrm{HdabcoClO_4}$.\cite{olejniczak_new_2013}}
    \label{fig:struct}
\end{figure}

Although many properties of plastic crystals have been characterized, a
microscopic understanding of the phase transitions and the nature of the disorder in plastic crystals is still largely missing. 
There is also limited insight into the polarization-switching mechanisms of these materials.\cite{shi_symmetry_2016} Such insight can be provided by molecular dynamics (MD) simulations;
however, parametrizing classical force fields can be non-trivial, particularly for systems 
with a complex bonding nature such as the hybrid ionic crystals.\cite{gonzalez_force_2011}
As the bonding picture can include charge transfer, highly anharmonic vibrations, hydrogen bonding, and in some cases proton transfer,
it may be hard to ensure that the specific functional form of the interaction potentials well describes all salient chemical effects of a systemvet. \cite{unke_machine_2021}
\textit{Ab initio} molecular dynamics, on the other hand, 
compute all electronic bonding effects, typically at the density functional theory (DFT) level.\cite{car_introduction_2002}
While this approach can provide much insight into smaller systems, computational costs can become prohibitive
for typical plastic crystals at the relevant time scales and supercell sizes.
The recent advent of machine-learned force fields (MLFFs) that can be 
trained on \textit{ab initio} data
has opened the door for predictive modeling of dynamic materials,
which with sufficient diverse data can approach the accuracy of the underlying DFT-based training data.\cite{wieser_machine_2024, unke_machine_2021, friederich_machine-learned_2021, carrete_deep_2023,wu_applications_2023,chmiela_machine_2017,schleder_dft_2019,fiedler_deep_2022}

In this work, we used such an approach to study phase transitions and dynamical properties of $\mathrm{HdabcoClO_4}$ in the 120-500 K range. 
Rather than a rotational switching mechanism, exhibited by many reported
ferroelectric plastic crystals,\cite{tang_visualization_2017, tang_organic_2020} $\mathrm{HdabcoClO_4}$
has a
displacive-type ferroelectric switching, which makes it capable of ferroelectric 
switching at frequencies up to $10~$kHz.\cite{tang_ultrafast_2016} 
It has a spontaneous polarization of $4.6~\mathrm{\mu C/cm^2}$ and a Curie temperature of $377~$K\cite{olejniczak_new_2013,li_improper_2021,katrusiak_ferroelectricity_1999} which is quite large for this class of compounds.  
The material also exhibits a rich phase diagram, dynamical disorder, and a partially orientationally disordered mesophase,  \cite{katrusiak_ferroelectricity_1999,katrusiak_proton_2000, olejniczak_new_2013, olejniczak_pressuretemperature_2018, li_improper_2021, tang_ultrafast_2016,olejniczak_pressuretemperature_2018} 
with nine phases reported below the decomposition temperature at $535~$K. \cite{olejniczak_new_2013} 
Fig. \ref{fig:struct} displays the structure of the paraelectric mesophase I, the room-temperature ferroelectric phase II, and the low-temperature ferroelectric phase III. 
In all the phases, the $\mathrm{Hdabco^{+}}$-molecules form hydrogen-bonded
columns. In the high-temperature phase I, partial proton disorder
has also been reported. 
Whereas the $\mathrm{ClO_{4}^{-}}$ anions have a large degree of oriental disorder in the mesophase, the $\mathrm{Hdabco^{+}}$-molecules predominantly show disorder around the hydrogen-bond direction.

\section{\label{sec:method}  Methods}

\subsection{Density functional theory}

The DFT calculations 
were based on the plane-augmented wave (PAW) formalism
\cite{blochl_projector_1994, kresse_ultrasoft_1999}
as implemented in the
Vienna Ab initio Simulation Package (VASP).\cite{kresse_ab_1994, kresse_ab_1993, kresse_efficiency_1996, kresse_efficient_1996}
The non-local van der Waals density functional
vdW-DF-cx~\cite{berland_review, berland_exchange_2014, facingchall} 
was selected as it can provide accurate lattice constants of highly diverse solids,\cite{Tran:choosing,facingchall,vdw_3} and we recently found it to provide accurate lattice constants of several plastic crystals.\cite{sodahl_piezoelectric_2023, seyedraofi_PT_CSD}
The DFT simulation cells were based on $2\times2\times2$ times the unit cell of phase II, corresponding to a cell 
with 416 atoms in total. 
In the DFT-MD simulations, the plane-wave cutoff was set to $530~$eV using a $\Gamma$-point sampling of the Brillouin zone. 
All DFT-MD simulations started from relaxed until cells. 
The DFT-MD simulations were carried out under the action of a Nóse-Hover thermostat\cite{nose_unified_1984,hoover_canonical_1985} and an NVT ensemble with a time step of $0.5~$fs.

\subsection{Machine-learned force field: training and simulations }
The machine-learned force field was trained using  the
neural-network-based
\textsc{NeuralIL}\cite{carrete_deep_2023} employing a Residual Neural network (ResNet) framework\cite{he_deep_2016} 
implemented on top of \textsc{Jax}\cite{jax2018github}  and  \textsc{Flax}.\cite{flax2020github}
In this method, the local environment within a radius of $r_{\mathrm{cut}}$ of an atom is decomposed into spherical Bessel descriptors.\cite{montes-campos_differentiable_2022} Using a local coordinate system ensures translational invariance, while rotational invariance is ensured by using the scalar power spectrum of the projections over basis functions. 
The chemical identity of the central atom is accounted for with embedding coefficients given by the type of element.
The descriptors and embedding coefficients are in turn fed into the neural network.\cite{carrete_deep_2023}
The core widths of the ResNet were set to 64:32:16. 
We used a batch size of eight and trained the MLFF for 25 epochs, which are sufficient thanks to the highly efficient nonlinear optimizer VeLO.\cite{metz_velo_2022}
In the training 
of the MLFF, we set the weight of the energy to 0.4 and the remaining 0.6 was assigned to forces.
A radial cutoff of $r_{\mathrm{cut}} = 4.0~\mathrm{Å}$ was selected, based on a convergence study increasing the $r_{\mathrm{cut}}$ from $3.5$ to $5.0~\mathrm{Å}$
in steps of 0.5~$\mathrm{Å}$, until we found no further reduction in error in the validation set. 
Similarly, the maximum radial order of the basis functions for the spherical Bessel descriptors was set to 4.
For the committee-based active learning, ten models were used to provide an uncertainty metric in the MLFF. 

The MLFF was trained in four stages. First, a crude model was trained using selected DFT-MD data. Second, we iteratively included additional DFT-MD data in six steps
by explicitly comparing the prediction error between the model and the DFT computations. 
Third, we used a committee-based active-learning procedure to obtain a more diverse training set and a stable MLFF. In the final step, we added training data in which the unit cells were compressed and expanded, as well as data where atomic species were swapped to alter the chemistry. This step provided further diversification of the training set and ensured that highly unfavorable configurations were represented in the data. 
$20\%$ of the configurations in the training set were randomly set aside for validation in each iteration of the training.

\begin{table}
\caption{\label{training_data} DFT-MD training data }
    \begin{tabular}{ccll}
         \hline
         \hline
          &Temperature &  Volume &   \\
         \hline
         A &$400$~K & $\mathrm{V_{relax}}$ &   \\
         B &$600$~K & $\mathrm{V_{relax}}$ &  \\
         C &$600$~K & $\mathrm{V_{relax}}$ & Forced mol. rot. \\
         D &$600$~K & $0.98\times \mathrm{V_{relax}}$ &  \\
         E &$600$~K & $1.02 \times \mathrm{V_{relax}}$ &   \\
         F &$800$~K & $\mathrm{V_{relax}}$ &  \\
         G &$800$~K & $0.9 \times \mathrm{V_{relax}}$  &  \\
        \hline
        \hline 
    \end{tabular}
\end{table}

\begin{table}
\caption{\label{tab:data_selection} Overview of MLFFs trained on selected DFT-MD data. $\mathrm{F_{RMSE}}$ and $\mathrm{E_{RMSE}}$ are the errors computed for the validation set. $\Delta \mathrm{F_{max}}$ is the maximum deviation between the MLFF and DFT-MD predicted forces.}
    \begin{tabular}{cccc}
         \hline
         \hline
          Configs. & $\mathrm{F_{RMSE}}$ (meV/Å) & $\mathrm{E_{RMSE}}$ (meV/atom) & $\Delta \mathrm{F_{max}}$ (eV/Å)\\
         \hline
          $400$ & $99$ & $3.0$ & $176$\\
          $450$ & $136$ & $5.9$ & $6.79$\\
          $500$ & $113$ & $7.3$ & $5.73$\\
          $550$ & $900$ & $4.9$ & $2.63$\\
          $600$ & $100$ & $8.3$ & $3.19$\\
          $650$ & $147$ & $5.4$ & $8.23$\\
          $700$ & $112$ & $4.6$ & \\
        \hline
        \hline 
    \end{tabular}
\end{table}

Initial DFT-MD simulations were carried out at different temperatures and volumes to obtain diverse yet physically representative starting training data, as shown in Table \ref{training_data}. 
For simulation C, one $\mathrm{ClO_{4}^{-}}$ and one $\mathrm{Hdabco^{+}}$ molecule were manually rotated in the initial configuration to force molecular rotations during the simulation. 
The first MLFF model was based on 400 configurations randomly selected from DFT-MD simulations C and D. 
Although this produced an MLFF with low root-mean-square errors (RMSE) for both forces and energies for the validation set, $99~$meV/Å and $3.0~$meV/atom, 
this model was inherently unstable, exhibiting cell “explosions” at $300~$K.
In the next step, $50$ configurations from all sets of DFT-MD data were added in each iteration
based on the largest deviations in force predictions between the MLFF and
the DFT-MD data, as shown in Table~\ref{tab:data_selection}.
Despite significantly reduced validation errors within the expanded training sets, 
subsequent MD simulations still resulted in unstable cell volumes for temperatures above $300~$K.
This illustrates that relying only on DFT-MD data to train an MLFF can be insufficient, as the short timescales feasible can be insufficient for providing sufficiently diverse training sets.
 
\begin{table}
\caption{\label{tab:active} Overview of MLFFs trained using active learning data and the errors computed from validation. $\mathrm{F_{RMSE}}$ and $\mathrm{E_{RMSE}}$ are the root square mean error in the forces on the configurations in the validation set. $\sigma_{\mathrm{max}}$ is the largest standard deviation computed in the active learning procedure.}
    \begin{tabular}{cccc}
         \hline
         \hline
          Configs. & $\mathrm{F_{RMSE}}$ (meV/Å) & $\mathrm{E_{RMSE}}$ (meV/atom) & $\sigma_{\mathrm{max}}$ (eV/Å) \\
         \hline
          $900$ & $ 100$ & $3.2$ & $121$\\
          $1100$ & $87$ & $5.5$ & $0.071$\\
          $1300$ & $114$ & $2.4$ & $0.005$\\
          $1500$ & $84$ & $3.0$ & $0.003$\\
        \hline
        \hline 
    \end{tabular}
\end{table}

In the committee-based active learning, ten MLFFs were trained using the same training set. 
Using the implementation of Carrete et al.,\cite{carrete_deep_2023}
all ten were trained in the same run with different initial random coefficients.
New atomic configurations were generated by running an MD simulation with a duration of $50~$ps using the MLFF model only trained on DFT-MD data. $1000$ configurations were evenly sampled and used as input for the committee. 
The standard deviation in the force predictions for the predictions of the committee was then used to identify atomic configurations that were not represented in the training set. 
Next, DFT computations were performed for the $200$ configurations with the largest standard deviations in forces, and the configurations were added to the training set. A new MLFF was then trained, and the procedure was repeated four times as shown in  Table \ref{tab:active}.  
The first three iterations used MLFF-MD simulations at $300~$K and $1~$bar. 
The fourth training set combined several MLFF-MD simulations at $400$ and $800~$K with pressures ranging from $1~$bar to $9~$kbar as input to the committee. After the fourth iteration, the volume predictions stabilized,  and the largest standard deviation in the volume was found for a simulation at $450~$K with a value of $1.6~\mathrm{Å^3}$/formula unit. 

Finally, the training data was further expanded to ensure that highly non-favorable configurations were represented in the training of the force field. We used two approaches to achieve this. $480$ configurations were constructed by scaling the unit cell parameters with a factor ranging from $0.9$ to $1.1$ in increments of $0.1$. This was applied for each unit cell parameter individually, but also to the volume of the cell. This results in compressed and expanded unit cells where the molecular geometries differ from their relaxed geometry. In addition, we constructed $200$ configurations in which two atoms in either $\mathrm{Hdabco^{+}}$ or $\mathrm{ClO_{4}^{-}}$ swapped positions. This ensured that less favorable chemistry was represented in the training data and can thus be appropriately avoided in the MD simulations. The forces and energies for all configurations were computed using DFT, and the final training set then contained $2180$ configurations. The resulting model had validation errors of $\mathrm{F_{RMSE} = 89~\mathrm{meV/Å}}$ and $\mathrm{E_{RMSE} = 4.5~\mathrm{meV/atom}}$.
 
MD simulations using the MLFF were performed using \textsc{Jax-MD},\cite{schoenholz_jax_2021} with a time step of $0.25~$fs. $30$ ps were used for thermalization, and the production runs were $180~$ps. An NPT ensemble was used with a Nosé-Hoover chain thermostat\cite{martyna_nosehoover_1992} and a barostat\cite{martyna_constant_1994} allowing flexible simulation cells  
using the integrator suggested by Yu et al.\cite{yu_measure-preserving_2010} as implemented by Bichelmaier et al.\cite{,bichelmaier_ab-initio_2023} 
The pressure was fixed at $1~$bar in all simulations. The simulations were initialized from simulation cells based on $6\times6\times6$ times of the unit cell of phase II of $\mathrm{HdabcoClO_4}$, which corresponds to a supercell size of $52.7\times 58.6 \times32.1~$Å, containing $11~232$ atoms or $432$ ionic pairs of $\mathrm{ClO_{4}^{-}}$ and $\mathrm{Hdabco^{+}}$. 
In total, 19 simulations with fixed temperatures in the  range between $120$ and $500~$K were performed
with a denser temperature sampling around the expected mesophase transition temperature.

\begin{figure}
    \centering
    \includegraphics[scale=0.50]{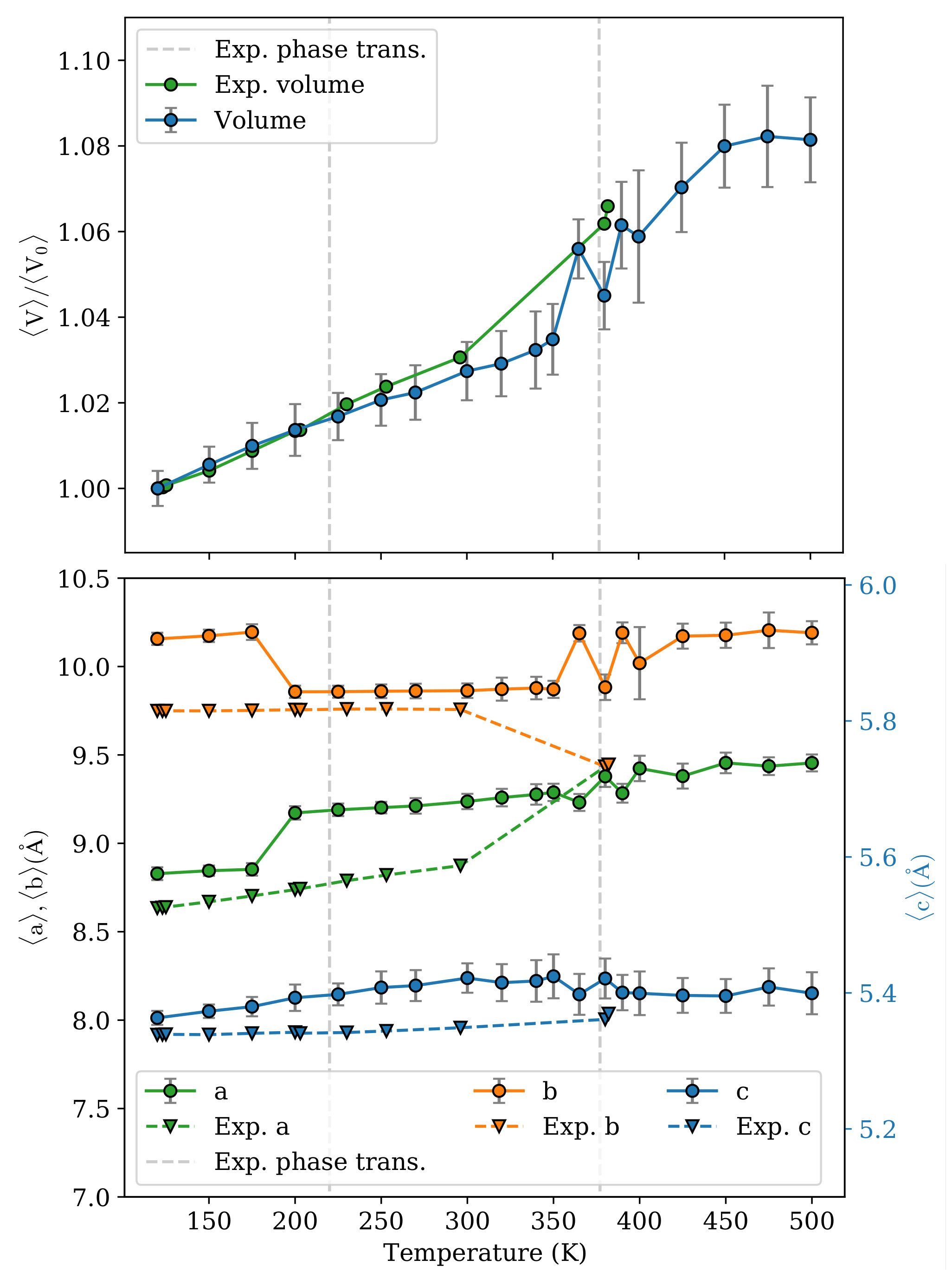}  \caption{ 
    Normalized experimental\cite{olejniczak_new_2013,olejniczak_pressuretemperature_2018,katrusiak_proton_2000} and predicted volumes (top), and experimental and predicted cell parameters (bottom). The bar half-length denotes two standard deviations. The cell parameters correspond to the phase-II lattice.
    The vertical dashed lines mark the experimental phase transition temperatures.\cite{olejniczak_new_2013}}
    \label{fig:expansion}
\end{figure}

\section{\label{sec:results}  Results and discussion}
In the following, we discuss the
thermal expansion, average displacement, and orientational disorder that arise in $\mathrm{HdabcoClO_4}$ at different temperatures. 

\subsection{Thermal expansion \label{sec:expansion}}

Fig. \ref{fig:expansion} plots the computed and experimental volumes normalized to those at $120~$K ($\mathrm{V_0}$) (top panel) and associated lattice constants (bottom). 
The computed volumes overestimate the experimental ones, by between $7\%$ for temperatures up to $300~$K and $5.5\%$ at $380~$K. 
The change in slope, i.e., the thermal expansion, from that below $350~$ K to that above $400~$ K, and the 
fluctuations in between is in line with the experimentally observed phase transition at  $377~$K. 

In the bottom panel of Fig.~\ref{fig:expansion}, the full lines indicate the computed lattice constants, given by $1/6$ of the supercell lattice parameters, which correspond to the lattice constants of phase II.
The hydrogen-bonded chains of $\mathrm{Hdabco^{+}}$ align with the $c$ axis. 
The computed values of $c$
agree well with experiment, with the largest deviation, an 
 overestimation of approximately $1.3~\%$, at $350~$K. 
For $a$ and $b$, the deviations are larger, up to $4.9~\%$ for both. The computed $a$ and $b$ values show anomalies at $200~$K, where $a$ increases and $b$ decreases, which is not reported experimentally.
At $380~$K, the experiment shows $a = b$, 
which is not found in our MD simulations, where $b$ instead exhibits a small, sudden increase.
The larger deviations for $a$ and $b$ 
may be due to limitations in MLFF and the training procedure, or it could be due to the choice of the exchange-correlation functional. Although vdW-DF-cx is highly accurate at typical equilibrium distances,\cite{sodahl_piezoelectric_2023} 
it tends to overestimate the interaction energies for dispersion-bonded molecular dimers beyond equilibrium.\cite{vdw_3,vdw_df_c6} 
This overestimation could lead to overestimated lattice constants in phases characterized by dynamic disorder. 
However, the experimental observation $a = b$ may also mask a more complex static or dynamic disorder occurring
at longer length scales and time scales than what can be probed with our MD simulation, but which is averaged out in the experimental characterization. \cite{morana_cubic_2023,krbal_2011,reuveni_static_2023}

\subsection{Ionic displacement and spontaneous polarization}

\begin{figure}
    \centering
    \includegraphics[scale=0.54]{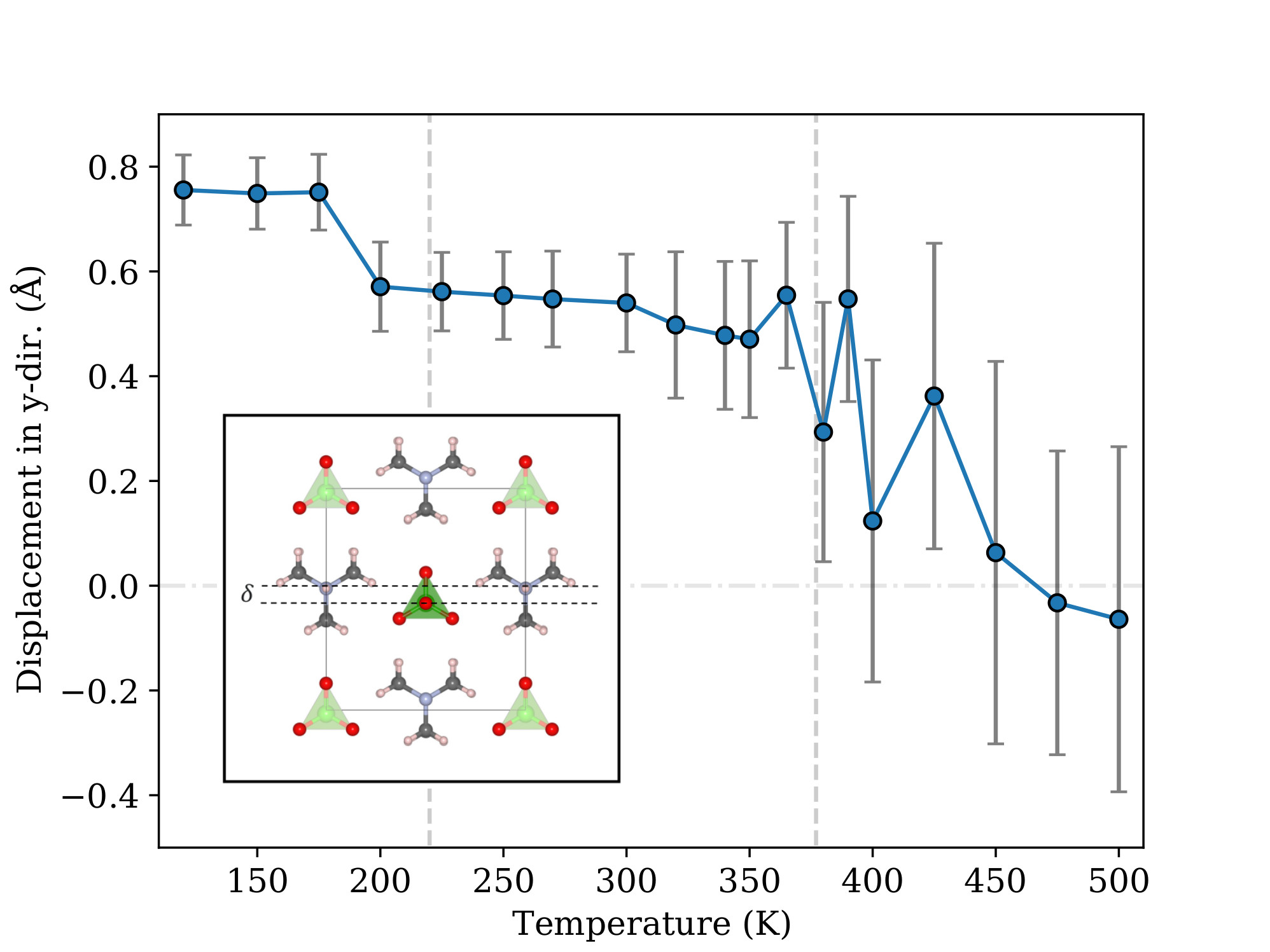}
    \caption{The average displacement $ \langle \delta \rangle $ of $\mathrm{Hdabco^{+}}$ molecules relative to the $\mathrm{ClO_{4}^{-}}$ molecules within the same layer in the simulation cell, with two standard deviations indicated. 
    The inset illustrates the displacement in phase II.}
    \label{fig:displacement}
\end{figure}

Fig. \ref{fig:displacement} shows the average ionic displacement $\delta$ in the $b$-direction
of $\mathrm{Hdabco^{+}}$ relative to the $\mathrm{ClO_{4}^{-}}$ columns. 
This parameter is linked to the spontaneous polarization of $\mathrm{HdabcoClO_4}$\cite{katrusiak_ferroelectricity_1999}
and serves as a ferroelectric-to-paraelectric order parameter. 
For temperatures up to $175~$K,  
$\delta \sim0.75 ~$Å,  before dropping to $0.58~$Å at $200~$K, coinciding with the transition between phases III and II.
At $380~$K there is also a marked drop in $\delta$ with a large increase in the corresponding deviations 
and further anomalous behavior before approaching low values beyond $450~$K, but with large deviations.
This is line with a broadened phase transition, where larger supercell sizes and/or longer time runs might result in sharp phase-transition temperatures.

\begin{figure*}
    \centering
    \includegraphics[scale=0.48]{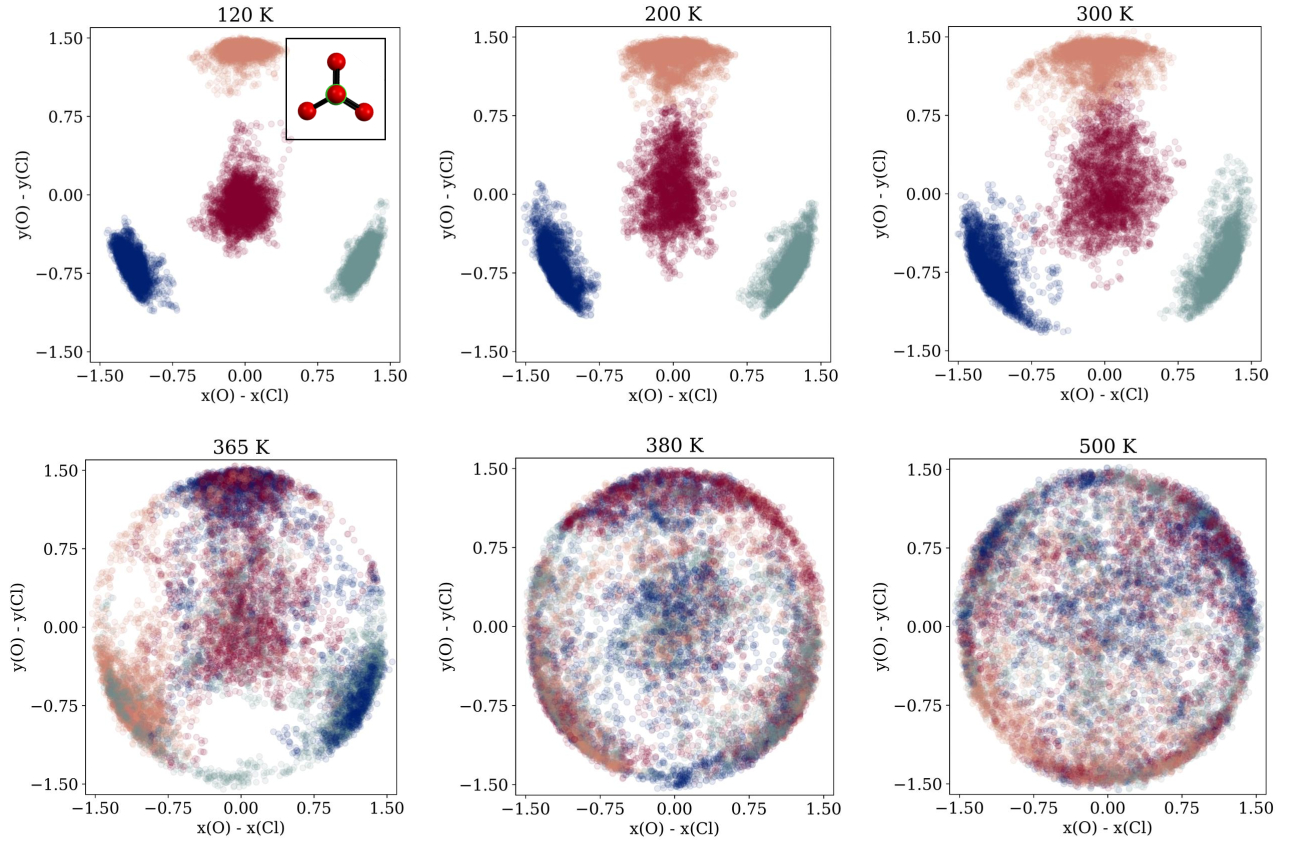}
    \caption{Oxygen atom positions relative to chlorine in a $\mathrm{ClO_4}$ throughout the simulation, colored by their initial position in the $\mathrm{ClO_4}$-molecule. The $\mathrm{ClO_4}$ is viewed from the above as indicated in the inset. 
    }
    \label{fig:ClO4_rot}
\end{figure*}

\begin{figure*}
    \centering \includegraphics[scale=0.45]{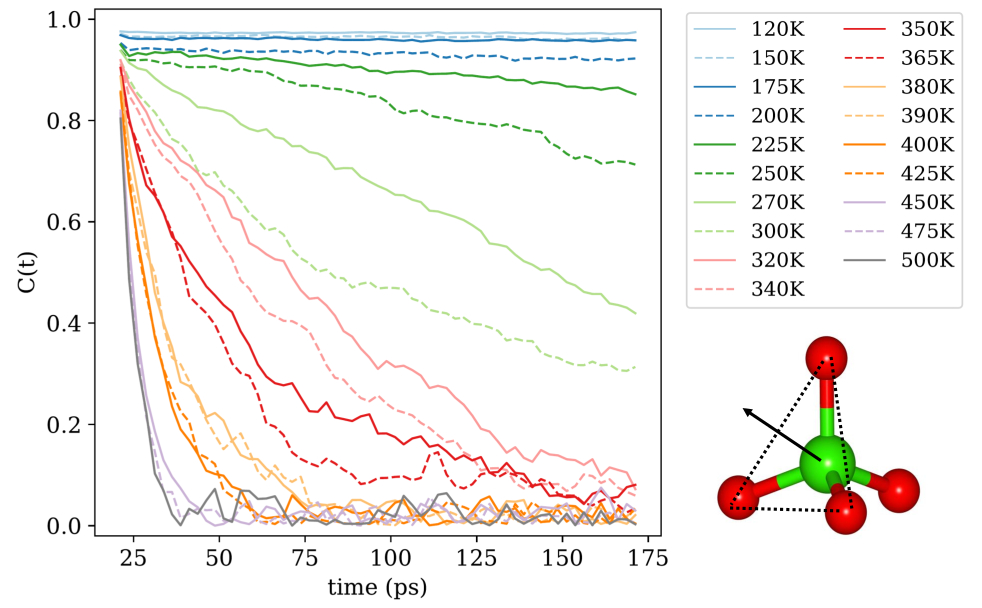}
    \caption{Rotational autocorrelation of $\mathrm{ClO_4}$ for different temperatures. The molecular direction for the autocorrelation is the vector from the chlorine atom to the center of a tetrahedral face, as illustrated to the right. }
    \label{fig:auto_cl}
\end{figure*}

\begin{figure*}
    \centering
    \includegraphics[scale=0.47]{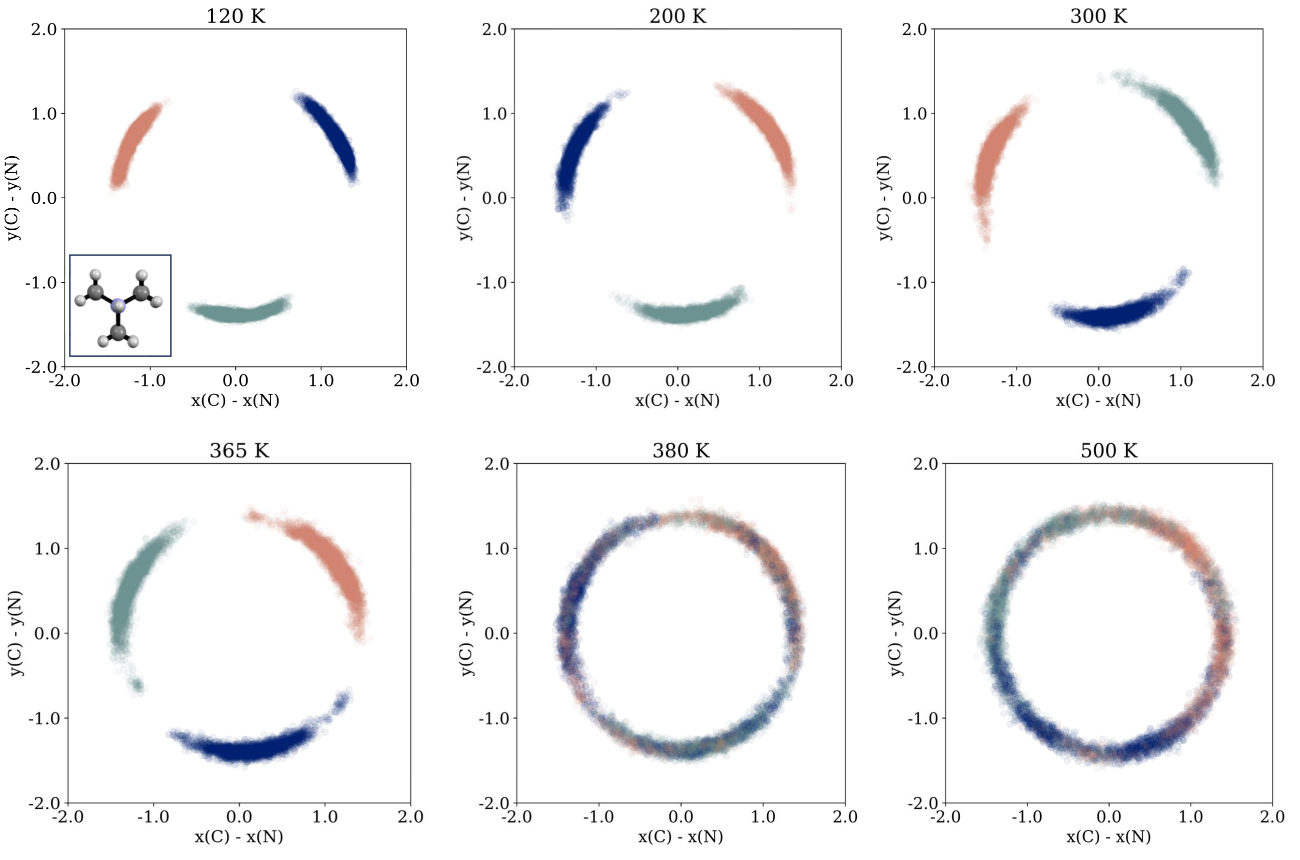}
    \caption{Illustration of the positions of three of the carbon atoms in a Hdabco-molecule relative to their nitrogen neighbor atom. The three colors each represent a distinct carbon atom in the molecule. The molecule is viewed along the hydrogen-bonded direction, as illustrated in the inset. The onset of molecular rotation around the hydrogen-bonded axis is at $380~$K. }
    \label{fig:dabco_rot}
\end{figure*}

\begin{figure*}
    \centering
    \includegraphics[scale=0.45]{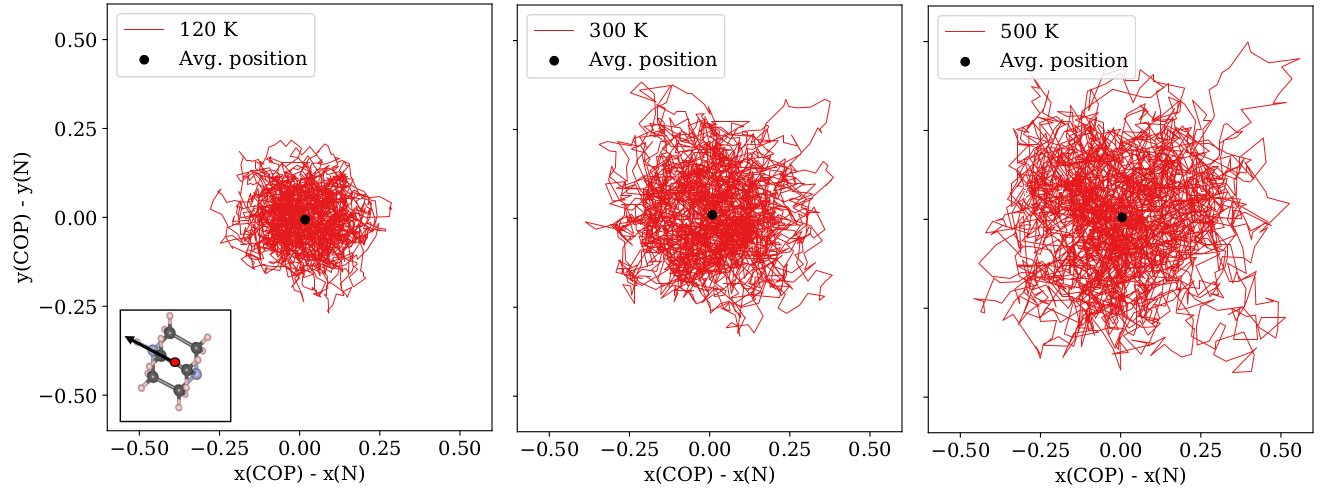}
    \caption{Position of the nitrogen atom in an $\mathrm{Hdabco^{+}}$-molecule relative to the center of the positions of the molecule, as illustrated in the inset (central position marked in red in the inset).}
    \label{fig:tilt}
\end{figure*}

\subsection{Orientational disorder}

\subsubsection{$\mathrm{ClO_{4}^{-}}$: Rotational dynamics}

Fig. \ref{fig:ClO4_rot} plots the oxygen atom positions (four different colors)
of a selected $\mathrm{ClO_{4}^{-}}$ anion in the x-y plane relative to its central Cl atom throughout an MD simulation for temperatures ranging from $120$ to $500~$K.  In the $120$-to-$300~$K range, the plots show that the tetrahedron has a preferred orientation, with libration motion that increases with temperature. At $365~$K, there is clearly a significant rotation as seen by the mixing of colors, but a distinctly preferred axis of orientation remains. 
This preference weakens at higher temperatures due to a transition into full rotational disorder in the x-y-plane, with also significant rotations on the sphere itself.  

The rotational disorder of the $\mathrm{ClO_{4}^{-}}$-molecules is evaluated using a rotational autocorrelation function \cite{yoneya_molecular_2020, adebahr_structure_2006}:

\begin{equation}
    C(t) = \frac{1}{N} \sum_i^{i=N} \mathbf{a}_{i} (t + t_0) \cdot \mathbf{a}_{i} (t)
\end{equation}
where $\mathbf{a}_{i}$ 
is given by a unit vector pointing from the central Cl atom to the center of a tetragonal face spanned by oxygen atoms.  
Fig. \ref{fig:auto_cl} plots $C(t)$ for temperatures between $120$ to $500~$K. 
$C(t)$ decreases with temperature but remains close to $0.9$ for temperatures between $120$ and $175~$K, i.e., an indication that no rotation occurs during the simulation. 
At $200~$K, $C(t)$ begins to steadily decrease, indicating the onset of occasional rotation of $\mathrm{ClO_{4}^{-}}$ anions, which increases with temperature. This finding is in line with a phase transition between phase III and II and the shift in $\delta$ found at this temperature in Fig.~\ref{fig:displacement}. 
The very rapid decay of $C(t)$ beyond $380~$K compared to the more conventional exponential decay at lower temperatures is also possibly reflecting the phase transition occurring between $365$ and $380~$K.

\subsubsection{$\mathrm{Hdabco^{+}}$: Rotation and tilting}

For the $\mathrm{Hdabco^{+}}$ molecules, we found rotation to only occur around the $c$ axis. 
Fig. \ref{fig:dabco_rot} plots the carbon atom position relative to the N-atom in the plane perpendicular to the c-axis, in the range of $120$ to $500~$K. 
Up to $365~$K, the plot shows increasing libration with temperature, 
but no onset of rotation.
At $365~$K, there is a larger spread in the carbon atom position, and at $380~$K and above, the trajectories indicate frequent rotations, in excellent agreement with the experimental phase transition temperature at $377~$K.

The constrained rotation of $\mathrm{Hdabco^{+}}$ molecules at elevated temperatures indicates that the hydrogen bonds are stable throughout the temperature range studied. This is also reflected in the volume expansion, as the length of $c$, 
the only hydrogen-bonded direction, 
is close to constant when temperature increases, even across phases. 
Hydrogen bonds have also been reported to be central to the mesophase behavior of plastic crystals. 
Yoneya and Harada\cite{yoneya_molecular_2020} studied quinuclidinium perrhenate using classical MD and found that a partially disordered phase was stabilized relative to the fully disordered mesophase, as intermolecular hydrogen bonds outcompete the thermal disorder for temperatures up to $367~$K. 

The onset of orientational disorder of $\mathrm{ClO_{4}^{-}}$ and $\mathrm{Hdabco^{+}}$ coincides with phases II and I, respectively. A similar behavior was reported for tetramethylammonium dicyanamide by Adebahr et al.\cite{adebahr_structure_2006}. They used MD with classical force fields and identified the onset of rotation of each of the two molecular entities as the driving mechanisms for two distinct phase transitions of the material

\begin{figure}
    \centering
    \includegraphics[scale=0.29]{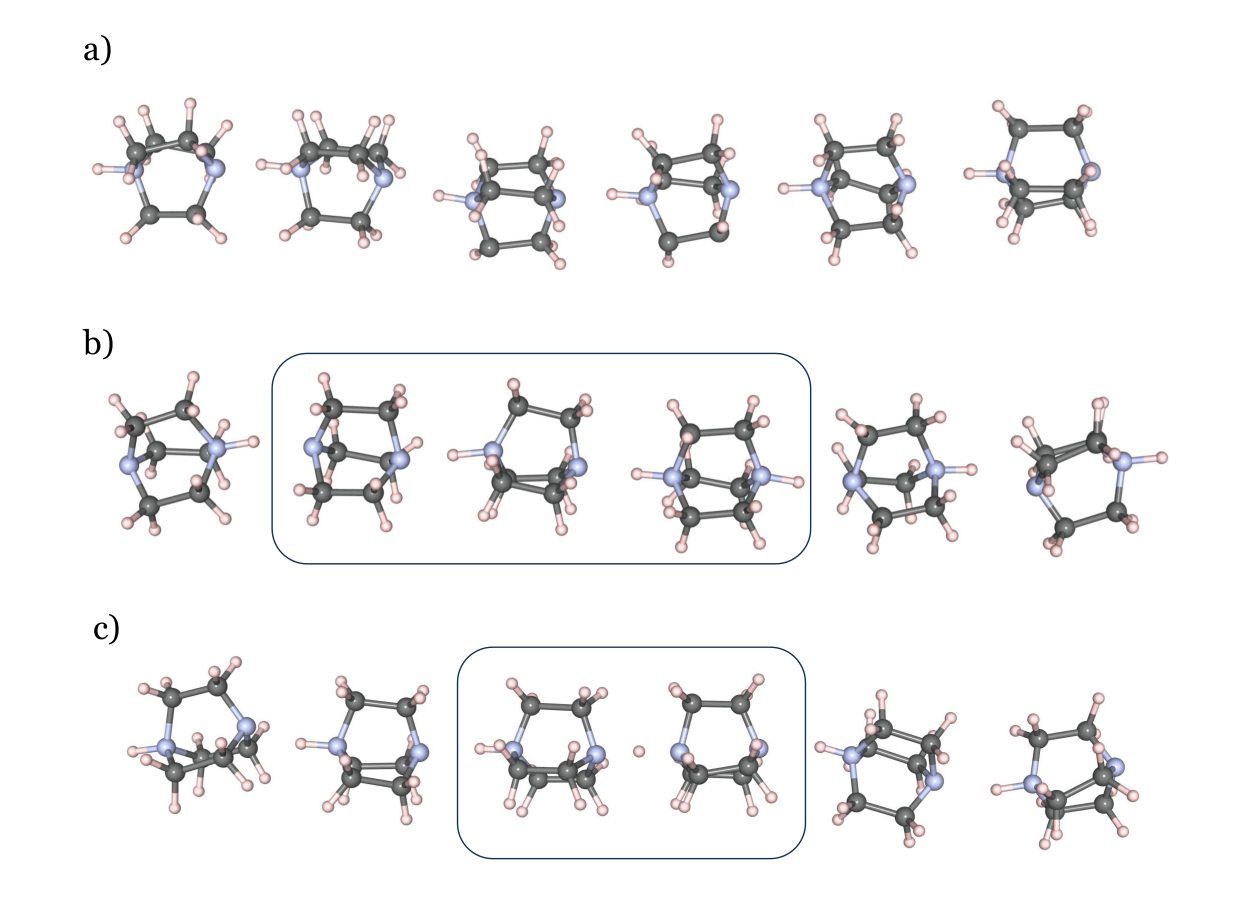}
    \caption{Snapshots of hydrogen-bonded chains at $425~$K: a) shows an aligned chain without defects, b) a chain with a double-protonated and deprotonated dabco-molecule, and c) a chain with a deprotonated molecule and a proton situated in the middle of two molecules.}
    \label{fig:proton_defects}
\end{figure}

Fig. \ref{fig:tilt} displays the position of one of the nitrogen atoms in a $\mathrm{Hdabco^{+}}$ molecule relative to the center of position of the molecule during simulation, as illustrated in the inset. The variation shows that in addition to the rotation around this center, the tilt of the cations increases with temperature.
Examples of the type of tilt the $\mathrm{Hdabco^{+}}$ cations exhibit relative to the direction of the chain
are also shown in Fig. \ref{fig:proton_defects}, obtained at $425~$K.

\subsection{Proton disorder and hydrogen bonds}

\begin{figure}
    \centering
    \includegraphics[scale=0.57]{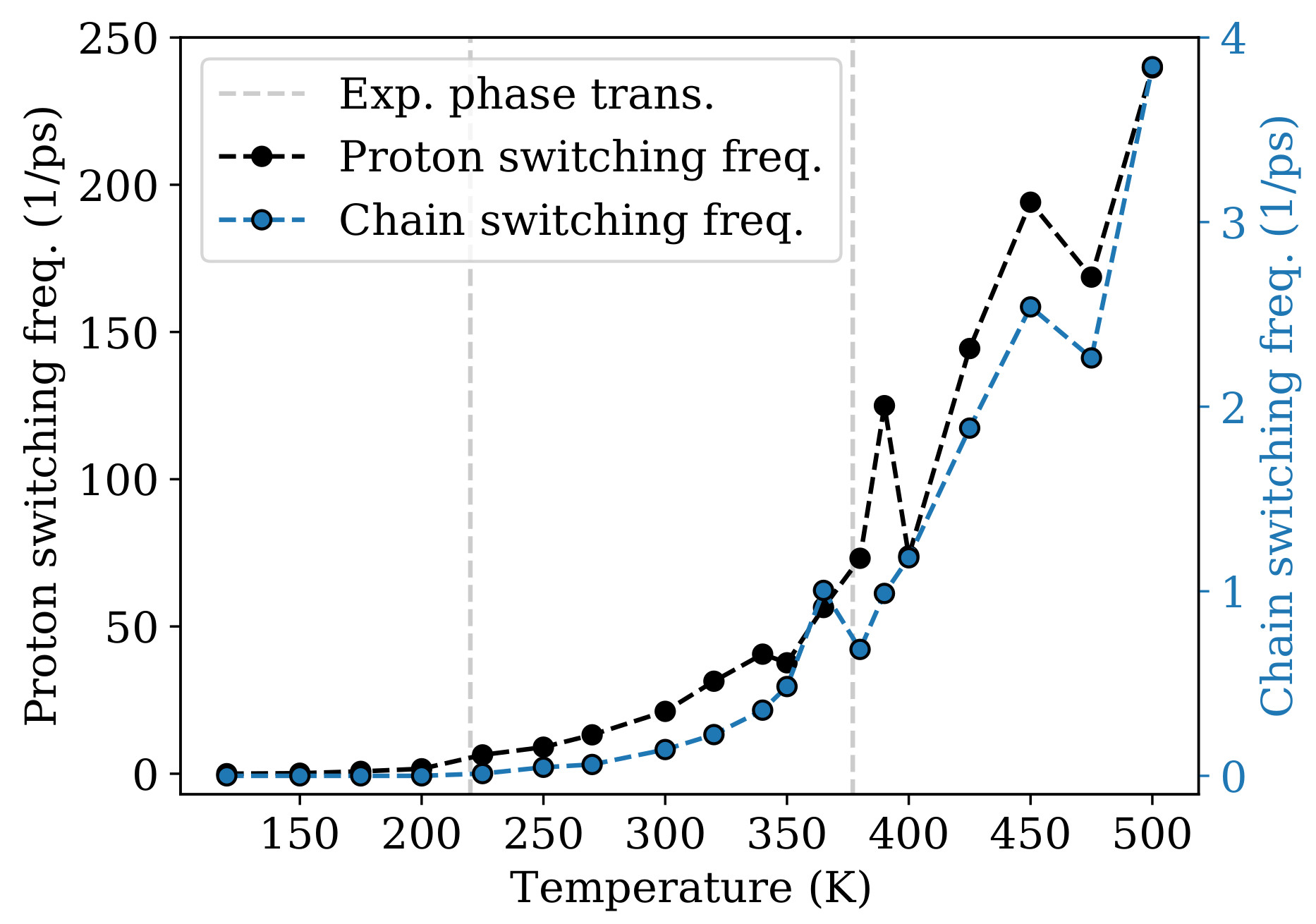}
    \caption{The frequency of proton transfer in $\mathrm{HdabcoClO_4}$. Protons are transferred at all temperatures, and the switching frequency of the orientation of hydrogen-bonded chains of $\mathrm{Hdabco^{+}}$-molecules.}
    \label{fig:frequencies}
\end{figure}

Proton disorder is also found in our simulations.
Fig.~\ref{fig:proton_defects} displays hydrogen-bonded chains of $\mathrm{Hdabco^{+}}$ molecules simulated at $425~$K. 
Fig. \ref{fig:proton_defects} 
a) illustrates a hydrogen-bonded chain of $\mathrm{Hdabco^{+}}$ molecules at $425~$K without defects, where all hydrogen bonds are oriented in the same direction. Fig. \ref{fig:proton_defects} b) shows a case where the transfer of a proton causes a defect where one cation is doubly protonated and another deprotonated. Fig. \ref{fig:proton_defects} c) illustrates a hydrogen-bonded chain where the proton is placed approximately in the middle of two cations. 

Such defects are observable already at temperatures of $150~$K and above. In Fig. \ref{fig:frequencies}, the black curve shows the frequencies of protons switching between two neighbor $\mathrm{Hdabco^{+}}$-molecules. The plot shows 
two changes in slope, one at $225$ and one around $365~$K. 
The blue curve shows the switching frequency of the orientation of the hydrogen-bonded chains. The trend is similar to the proton transfer frequencies, but the switching frequency of a whole chain is two orders of magnitude lower. This shows that most proton transfer events create short-lived local defects.

\begin{figure}
    \centering
    \includegraphics[scale=0.19]{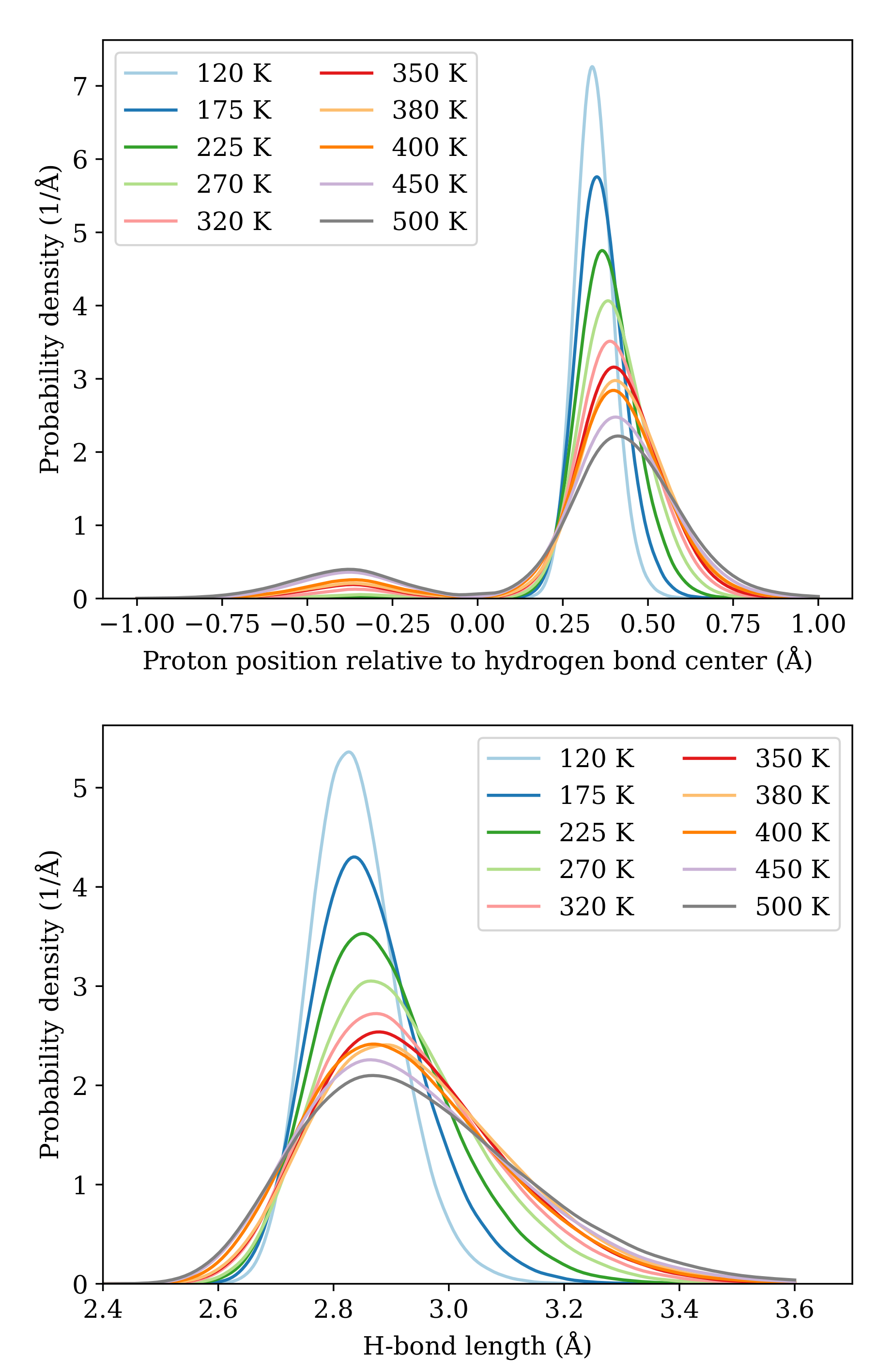}
    \caption{The probability density of the proton positions relative to the center of their hydrogen bond (top), and the corresponding hydrogen bond lengths between $\mathrm{Hdabco^{+}}$-molecules (bottom). 
    Positive proton position values indicate that the proton is oriented in the same as the hydrogen-bonded chain of $\mathrm{Hdabco^{+}}$-molecules, while negative indicates the opposite.
    \label{fig:prob_densities}}
\end{figure}
Fig. \ref{fig:prob_densities} (top panel) plots the proton position relative to the middle of its hydrogen bond. Here, positive values reflect protons in hydrogen bonds
that are oriented in the same direction as the overall orientation of its hydrogen-bonded chain. Negative values indicate that the proton is in a hydrogen bond with an opposite orientation relative to the chain. The plot shows a higher probability of finding protons aligned with the chain than ones that do not align with the chain directions at all temperatures
This preference shows that there is still a directionality of the hydrogen-bonded chains at elevated temperatures and not a full disorder of protons. 
Similar, in the bottom panel of Fig. \ref{fig:prob_densities} shows the distribution of the lengths of hydrogen bonds between $\mathrm{Hdabco^{+}}$ molecules,
showing a slight  increase in most typical bond lengths, but also larger fluctuations in the bond lengths as temperature increases. 

While this study provides qualitative insight into proton transfer in organic systems that may carry over to other organic and hybrid crystals, DFT computations of proton transfer barriers are very sensitive to the exchange-correlation functional employed. Seyedraoufi and Berland\cite{seyedraoufi_improved_2022} recently found 
for a set of molecular dimers that while vdW-DF correlation
can significantly improve proton transfer barriers compared to 
using correlation at the generalized gradient approximation (GGA), which severely underestimated barrier heights,
the vdW-DF-cx variant also underestimates such barriers, due to its "soft" exchange form,\cite{facingchall,Jenkins_2021} 
Using vdW-DF2,\cite{lee_higher-accuracy_2010} which predicted more accurate barriers could thus have improved the accuracy for this system, and so would adopting a hybrid functional, such as the vdW-DF-cx0-20 functional.~\cite{berland_assessment_2017,jiao_extent_2018}
Such a functional would have likely significantly delayed the onset of proton transfer.

\subsection{Summary of phase transitions parameters}

\begin{table}
    \centering
    \begin{tabular}{l|cc}
    \hline\hline
         & III to II (K) & II to I (K)\\
         \hline
       Experimental phase trans. & $220$ & $377$  \\
       \hline
       Volume expansion & -- & $380$ \\
       Ferro. displacement & $200$ & $380-450$ \\
       $\mathrm{ClO_{4}^{-}}$ disorder & $225$ & $380$ \\
       $\mathrm{Hdabco^{+}}$ disorder & -- & $380$ \\
       H-bond chain switching & $225$ & $365$ \\ 
     \hline
     \hline
    \end{tabular}
    \caption{Experimental phase transition temperatures\cite{olejniczak_new_2013} and an overview of the onset of disorder and symmetry changes in the MLFF simulations of $\mathrm{HdabcoClO_4}$.}
    \label{tab:Hdabco_PTs}
\end{table}

A summary of computed phase transition characteristics of
$\mathrm{HdabcoClO_4}$ are listed in Tab. \ref{tab:Hdabco_PTs},
alongside the experimental phase transition temperatures.\cite{olejniczak_new_2013} 
The computational results are consistent with experimental measurements overall. 
In the transition from III-II, neither theory nor experiment found any marked changes in volume; although in the computations, we found changes in the lattice constants as seen in Fig.~\ref{fig:expansion}.
The reduction in displacement of $\mathrm{Hdabco^{+}}$ molecules relative to $\mathrm{ClO_{4}^{-}}$ molecules at $200~$K,
as shown in Fig.~\ref{fig:displacement} is also indicative of a phase transition between these two ferroelec2tric phases,
and so is the onset of orientation disorder between $200$ and $225~$ K, as evidenced by Figs.~\ref{fig:ClO4_rot} and \ref{fig:auto_cl}.
Moreover, the MD data in Figs.~\ref{fig:ClO4_rot} and \ref{fig:dabco_rot} 
is in line with the increased thermal vibration reported for this phase.\cite{olejniczak_pressuretemperature_2018}
The clear slope change in line with proton disorder in phase II at $200~$K, is also in line with a phase transition, but experimentally proton disorder has only been observed in phase III.\cite{olejniczak_pressuretemperature_2018}

For the phase transition between II and I, similar features to those in the experiment are found,
albeit spread at different temperatures. 
The autocorrelation of $\mathrm{ClO_{4}^{-}}$ (Fig.~\ref{fig:auto_cl} )
indicates the onset of essentially free rotation of this species at $380~$K, in line with experiment. 
Moreover, at this temperature, we also find the onset of molecular rotations of 
$\mathrm{Hdabco^{+}}$. 
The change in sublattice displacement (Fig.~\ref{fig:displacement}) is less clear, showing increasing deviations from temperatures of $380~$ K and above,
while reaching values close to zero first at $450~$K. 
This apparent disparity with experiments may also hint at nanoscale domain formation,
i.e., using much larger supercells and longer time runs might average out to provide a cubic unit cell in line with experiment structure characterization

\section{\label{sec:discuss} Conclusion and outlook}

An MLFF was trained for $\mathrm{HdabcoClO_4}$ using the neural network \textsc{NeuralIL},
with an active learning procedure to diversify the training set.
Our study highlights how MLFF-based MD can be used to gain fundamental insight into the dynamical properties of plastic ionic crystals, with overall encouraging agreement between computed and measured phase transition properties. 
By using a fully ab initio approach that requires no knowledge of predefined bonding properties, our study highlights
how MLFF can be used both for computational design and analysis
of the emerging class of dynamical materials such as plastic ionic crystals, 
in particular in combination with advanced structural characterization methods. 
However, the disparities between theory and experiment also highlight the need for systematic benchmarking of both MLFF approaches and DFT exchange-correlation functionals for out-of-equilibrium geometries for systems exhibiting complex non-covalent bonding, such as plastic ionic crystals.    

\section*{Data availability}
All training data can be accessed through the Nomad database with DOI:10.17172/NOMAD/2024.09.19-1.

\section*{\label{sec:acknowledge} Acknowledgements}
We thank O. Nilsen, C.H. Gørbitz, R. Tranås, and H. H. Klemetsdal for valuable discussions. 
Work by EDS and KB supported by the Research Council of Norway as a part of the Young Research Talent project FOX (302362).
The computations were carried out on UNINETT Sigma2 high-performance computing resources (grant NN9650K). This study was supported by MCIN with funding from the European Union NextGenerationEU (PRTR-C17.I1) promoted by the Government of Aragon. J.C. acknowledges funding from MICIU/AEI (DOI:10.13039/501100011033) through grant CEX2023-001286-S.

\bibliography{references}





\end{document}